\documentclass[conference,a4paper]{APSIPA2025}
\usepackage{amsmath}
\usepackage{svg}
\usepackage{graphicx}
\usepackage{multirow}
\usepackage{threeparttable}
\usepackage[backend=biber,style=ieee]{biblatex}
\usepackage{geometry}
\usepackage{fancyhdr}

  \fancypagestyle{firststyle}{
    \fancyhf{}
    \fancyhead[C]{2025 Asia Pacific Signal and Information Processing Association Annual Summit and Conference (APSIPA ASC)}
  }

\begin{document}

\title{Guitar Tone Morphing by Diffusion-based Model}
\author{
  \authorblockN{
    Kuan‐Yu Chen\textsuperscript{1*}, 
    Kuan‐Lin Chen\textsuperscript{1*}, 
    Yu‐Chieh Yu\textsuperscript{1*},
    Jian‐Jiun Ding\textsuperscript{1}
  }
  \authorblockA{
    \textsuperscript{1}Graduate Institute of Communication Engineering, National Taiwan University, Taiwan\\
    \textsuperscript{*}These three authors contributed equally.\\
    E‐mail: \{r13942135, r13942067, r12942142, jjding\}@ntu.edu.tw
  }
}
\maketitle
\thispagestyle{firststyle}
\pagestyle{fancy}

\begin{abstract}
In Music Information Retrieval (MIR), modeling and transforming the tone of musical instruments, particularly electric guitars, has gained increasing attention due to the richness of the instrument tone and the flexibility of expression. Tone morphing enables smooth transitions between different guitar sounds, giving musicians greater freedom to explore new textures and personalize their performances. This study explores learning-based approaches for guitar tone morphing, beginning with LoRA fine-tuning to improve the model performance on limited data. Moreover, we introduce a simpler method, named spherical interpolation using Music2Latent. It yields significantly better results than the more complex fine-tuning approach. Experiments show that the proposed architecture generates smoother and more natural tone transitions, making it a practical and efficient tool for music production and real-time audio effects.
\end{abstract}

\section{Introduction}

Guitar tone morphing is to perform transformation from one guitar sound to another smoothly, enabling more expressive performance and creative exploration. While traditional tone shaping relies on discrete effects or preset changes, morphing offers continuous control over timbre, bridging tonal spaces that were previously separated.

Early approaches to tone morphing were based on signal processing. Dudley’s Vocoder \cite{dudley1939voder} decomposed audio into frequency bands for real-time transformation. The phase vocoder \cite{flanagan1966phase} improved on this by using short-time Fourier transforms to enable time-stretching and pitch-shifting. Additive resynthesis \cite{serra1990spectral} constructed tone flexibly by controlling sine wave parameters.

With the rise of deep learning, guitar tone modeling has become increasingly data-driven. NSynth \cite{engel2017neural} used a WaveNet autoencoder to interpolate between instruments in latent space. GANSynth \cite{donahue2019gansynth} improved audio fidelity by jointly modeling magnitude and phase.

Recently, diffusion models have shown strong performance in audio generation. DiffWave \cite{kong2021diffwave} produced high-quality waveforms through iterative denoising, while AudioLDM \cite{liu2023audioldm} and MusicLDM \cite{chen2024musicldm} extended these techniques to music datasets and text-to-audio tasks. Style interpolation has been explored through LoRA fine-tuning in SoundMorpher \cite{niu2024soundmorpher} and QKV manipulation for controllable morphing \cite{10890164}. In parallel, Music2Latent \cite{pasini2024music2latent} offered a lightweight, content-preserving approach to audio interpolation without diffusion.

Visual-domain methods such as IMPUS \cite{yang2024impus} and DiffMorpher \cite{zhang2024diffmorpher} have also influenced our approach. These models achieve perceptual uniformity by interpolating both latent vectors and network weights.

In this work, we explore guitar tone morphing through four methods: three are based on latent diffusion models with various LoRA configurations, and one is to apply direct latent interpolation in Music2Latent. By combining ideas from the neural style transfer, latent blending, and diffusion generation, smooth and perceptually natural transitions between guitar tones with shared musical content can be achieved.
\section{Related Work}

Research on direct audio tone morphing remains limited, particularly for guitar tones. Existing studies mostly explored switching between distinct effects rather than smooth transitions. MorphFader \cite{10890164} used a TTAs model to blend two sounds by adjusting the attention mechanisms and was able to control extreme tone changes. SoundMorpher \cite{niu2024soundmorpher} applied a diffusion-based approach to generate perceptually smooth audio transitions, emphasizing uniform and natural-sounding morphs.

While audio tone morphing is still emerging, computer vision (CV) has advanced significantly in morphing and interpolation, providing valuable insights. For example, IMPUS \cite{yang2024impus} used diffusion to generate perceptually uniform image transitions. DiffMorpher \cite{zhang2024diffmorpher} further improved this by interpolating both LoRA weights and noise input, allowing label-free and fine-grained visual morphing. The Music2Latent framework \cite{pasini2024music2latent} provided useful guidance on style-content disentanglement in audio. These techniques inspired us to develop an advanced guitar tone interpolation approach.

In this work, guitar tone morphing is formulated as interpolation between mel spectrograms. Using latent diffusion models (LDM) \cite{rombach2022high}, we adapt image interpolation methods to smoothly transition between audio tones, even when the inputs are significantly different.

We also apply the technique of audio style transfer, which transforms the sound style (e.g., through different amps) while preserving content. 
Since our goal is to interpolate rather than transfer, maintaining content stability while blending tonal characteristics is an important issue. 

In summary, our approach adapts diffusion-based morphing, originally from image processing, to guitar tone interpolation. This enables smoother and more natural transitions between tones. We also integrate audio style transfer to enhance flexibility and expressiveness, making the system both natural to the ear and more controllable for users shaping timbre to their artistic goals.

\begin{figure*}[t]
  \centering
  \includegraphics[width=\textwidth]{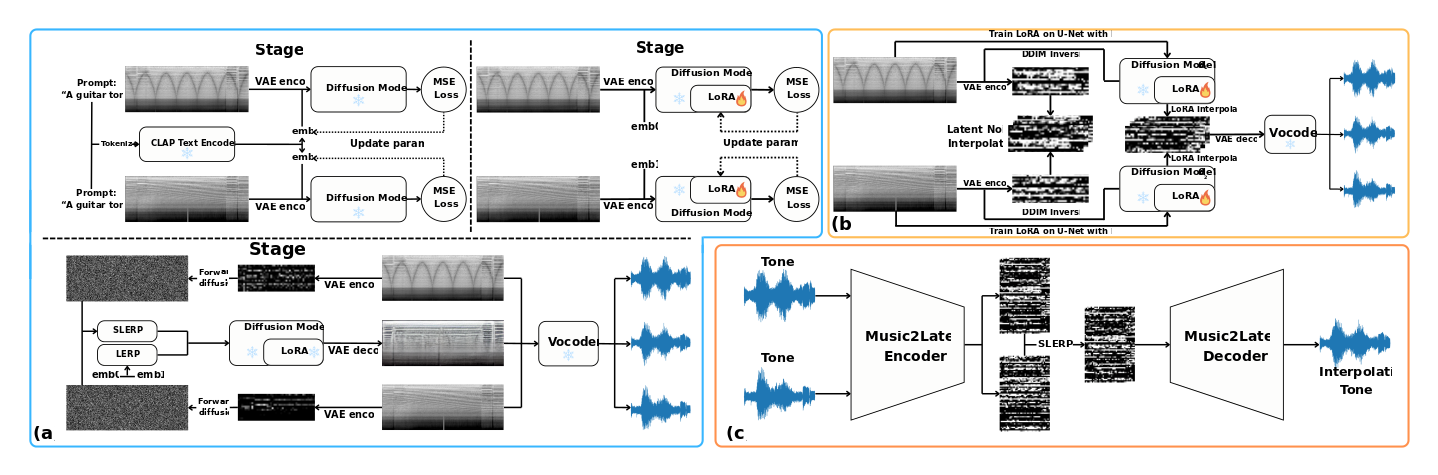} 
  \caption{Overall architecture of the proposed methods: (a) Single-Sided LoRA fine-tuning applied to the decoder; (b) Dual-Sided LoRA applied to both encoder and decoder; (c) Music2Latent interpolation that directly performs spherical interpolation in the latent space without parameter updating.}
  \label{fig:model}
\end{figure*}

\section{Methodology}

In this study, we propose four different methods for guitar tone interpolation. The first three methods are based on the Latent Diffusion Model (LDM)\cite{liu2023audioldm, liu2024audioldm, chen2024musicldm} and differ mainly in how they utilize LoRA fine-tuning and whether they rely on textual prompts or purely latent representations. The fourth method uses the Music2Latent framework \cite{pasini2024music2latent}, which does not use diffusion models and instead directly interpolates within a learned latent space. We detail each of the four proposed approaches individually.

\subsection{Interpolation and Style Transfer Techniques}

Several important interpolation and style transfer techniques are utilized in the proposed approaches, including Spherical Linear Interpolation (SLERP), Linear Interpolation (LERP), and Adaptive Instance Normalization (AdaIN). 

\paragraph{Spherical Linear Interpolation (SLERP)}

SLERP provides smooth interpolation between two vectors on the unit sphere, ensuring geometrically consistent transitions. Given two vectors $\mathbf{v}_0$ and $\mathbf{v}_1$, SLERP is performed as follows. 

First, both $\mathbf{v}_0$ and $\mathbf{v}_1$ are normalized:
\begin{equation}
\hat{\mathbf{v}} = \frac{\mathbf{v}}{\|\mathbf{v}\|}.
\label{eq:normalize}
\end{equation}
Next, we compute the angle $\theta_0$ between them:
\begin{equation}
\theta_0 = \arccos{(\hat{\mathbf{v}}_0 \cdot \hat{\mathbf{v}}_1)}.
\label{eq:angle}
\end{equation}
When the vectors are nearly identical (i.e., $\theta_0$ is very small), linear interpolation (LERP) is used for interpolation instead to avoid numerical problems.
\begin{equation}
\text{LERP}(\alpha, \hat{\mathbf{v}}_0, \hat{\mathbf{v}}_1) = (1 - \alpha)\hat{\mathbf{v}}_0 + \alpha \hat{\mathbf{v}}_1.
\label{eq:lerp}
\end{equation}
Otherwise, SLERP is computed by:
\begin{equation}
\text{SLERP}(\alpha, \hat{\mathbf{v}}_0, \hat{\mathbf{v}}_1) = 
\frac{\sin((1 - \alpha)\theta_0)}{\sin(\theta_0)} \hat{\mathbf{v}}_0 + 
\frac{\sin(\alpha \theta_0)}{\sin(\theta_0)} \hat{\mathbf{v}}_1.
\label{eq:slerp}
\end{equation}

\paragraph{Adaptive Instance Normalization (AdaIN)}

AdaIN \cite{huang2017arbitrary} is applicable to style transfer tasks. It adjusts the mean and variance of one feature map ($x$) to match another ($y$), thus aligning the style distributions:

\begin{equation}
\text{AdaIN}(x, y) = \sigma(y) \left( \frac{x - \mu(x)}{\sigma(x)} \right) + \mu(y).
\label{eq:adain}
\end{equation}

\subsection{Without LoRA Fine-tuning}

This baseline method does not incorporate any LoRA fine-tuning. We begin by applying \textbf{Textual Inversion} to enhance the expressiveness of prompts in the latent space. Then, we encode two input audio clips using the pre-trained LDM encoder and directly interpolate their latent vectors using SLERP. The interpolated vector is decoded back into the Mel-spectrogram by the original decoder and is finally transformed into a waveform using a vocoder. This approach allows us to evaluate the tone interpolation capabilities of a purely pre-trained diffusion model without any task-specific adaptation.

\subsection{Single-Sided LoRA Fine-tuning}

Building on the AudioLDM framework \cite{liu2023audioldm}, this method introduces LoRA fine-tuning to enhance prompt control and decoding quality. After performing textual inversion, we fine-tune only the conditional U-Net using LoRA to align text embeddings with latent features. Meanwhile, the unconditional U-Net is fine-tuned independently with a LoRA rank of 2, without textual input. During inference, we apply SLERP to interpolate latent vectors from the two input tones and perform linear interpolation (LERP) on their text embeddings. The interpolated representation is then decoded through the unconditional U-Net, the VAE, and the vocoder. This method enables more expressive tone blending by partially adapting the model to the task.

As illustrated in Fig.~\ref{fig:model}(a), the LoRA fine-tuning is applied only to the decoder side (U-Net), while the encoder remains frozen.

\subsection{Dual-Sided LoRA Fine-Tuning}

This method extends the single-sided approach by fine-tuning two separate U-Net models. Different models are applied for different input tones. Following textual inversion, each U-Net is individually fine-tuned using LoRA with rank 2 to specialize in its respective tone domain. During inference, we perform SLERP on the latent vectors and LERP on the prompt embeddings, as in the previous method. Additionally, we interpolate the parameters of the two U-Nets via LERP to construct a merged model tailored to the interpolated tone. The final latent data are decoded using the merged U-Net, followed by the VAE and the vocoder.

To further refine tone characteristics, we integrate AdaIN into the decoding pipeline. AdaIN makes the mean and variance of the output match those of the interpolated tone style, allowing for more precise control over the stylistic attributes of the output tone.

As shown in Figure~\ref{fig:model}(b), both conditional and unconditional U-Nets are fine-tuned independently and their parameters are interpolated at the inference process.

\subsection{Music2Latent Interpolation}

This method employs the Music2Latent framework \cite{pasini2024music2latent}, which does not rely on diffusion models or text conditioning. Instead, it uses a lightweight encoder–decoder architecture for latent audio reconstruction. We encode both input audio signals, interpolate their latent vectors using SLERP, and decode the result back into audio. While lacking fine-tuning, this method offers a clean benchmark for evaluating the interpolation quality within a learned latent space. We also apply AdaIN to better align the style characteristics during latent morphing, enhancing the naturalness of the resulting tone.

As illustrated in Figure~\ref{fig:model}(c), the system directly encodes and decodes the audio signal without text prompts or parameter tuning, relying purely on latent-space interpolation.
\section{Experiments and Results}

\subsection{Dataset}

Our dataset consists of real guitar recordings that were processed using a variety of tone-shaping effects to simulate a wide range of guitar tones. This approach ensures diversity and realism in the tonal characteristics captured. The dataset contains a wide range of professionally recorded tones that well reflect real-world playing conditions.

We define \textit{five core tone morphing tasks}, each corresponding to a commonly used guitar tone transition inspired by standard guitar effects \cite{brewster2003introduction}. Each task contains 20 pairs of audio clips, where each pair represents a source-target tone mapping for evaluation or fine-tuning. The tasks are:

\begin{itemize}
\item \textbf{Clean to High Gain:} transforming a clean tone into one with heavy distortion.
\item \textbf{Clean to Low Gain:} adding light distortion to a clean tone.
\item \textbf{Low Gain to High Gain:} increasing distortion on an already mildly distorted tone.
\item \textbf{Clean to Modulation:} applying effects like chorus or flanger to a clean tone.
\item \textbf{Modulation to High Gain:} Combining modulation effects with heavy distortion.
\end{itemize}

Each pair consists of two short clips (about 5 seconds each). While the original recordings were sampled at the rate of 44.1 kHz, most diffusion-based models for comparison, such as AudioLDM and SoundMorpher, require the audio to be downsampled to 16 kHz due to model constraints. As a result, we performed most experiments using the 16 kHz version of the dataset. However, the proposed model of spherical interpolation using Music2Latent supports native 44.1 kHz input, allowing it to preserve more detailed high-frequency content, which gives it a potential advantage in \textbf{timbral accuracy} and \textbf{clarity} during tone morphing.

Each experiment was conducted at a local pairwise level, where fine-tuning or interpolation is directly applied to each input-output tone pair. This setup reflects real-world tone morphing situations. As shown in Fig.~\ref{fig:model}, fine-tuning is only performed on the two input sides in these cases.

\subsection{Evaluation Metrics}

\subsubsection{Interpolation Tone Quality Assessment Metrics}

To comprehensively evaluate the quality of tone morphing, we employ both objective and subjective metrics:

\begin{itemize}
\item \textbf{CDPAM mean ± std} is a perceptual audio similarity metric derived from deep models. It reflects how consistently the generated audio is perceived as similar to the target, based on features aligned with human hearing. The mean indicates average similarity, while the standard deviation captures variability across different comparisons.
\item \textbf{Mean Opinion Score (MOS)} provides a subjective evaluation of audio quality. In this study, 20 human raters participated in a blind test to rate the naturalness and smoothness of the morphed outputs on a scale from 1 (poor) to 5 (excellent).
\end{itemize}

The objective metric CDPAM offers a quantitative and reproducible way to assess audio similarity, while MOS complements them by capturing perceptual nuances that automated metrics might overlook. Notably, MOS remains a widely trusted indicator in audio generation research, making it a critical component of our evaluation protocol. By combining these metrics, we gain a well-rounded understanding of how effective each model is in producing high-quality, perceptually convincing tone morphs.

\subsubsection{Spectral Convergence (SC) for Reconstruction Quality Evaluation}

To assess the reconstruction quality of the generated audio signals—particularly how faithfully the models decode internal representations back into waveform—we adopt the \textbf{Spectral Convergence (SC)} loss, a key component of the multi-resolution Short-Time Fourier Transform (STFT) loss.

This metric directly evaluates differences in the frequency domain by comparing the spectrograms of the predicted and ground truth waveforms. Let $M_r$ and $M_t$ denote the magnitude spectrograms of the reconstructed and target audio, respectively. The SC loss is formulated as:

\begin{equation}
\text{SC}(M_r, M_t) =
\frac{
\sqrt{\sum_{m,k} \left(M_t(m,k) - M_r(m,k)\right)^2}
}{
\sqrt{\sum_{m,k} M_t(m,k)^2}
}.
\label{eq:spectral_convergence}
\end{equation}

Here, $m$ and $k$ represent the time and frequency indices in the spectrogram. Intuitively, the SC loss measures the normalized deviation between the predicted and target spectrograms, where a lower value indicates better spectral reconstruction.

To capture both short- and long-term frequency patterns, we compute SC values across multiple STFT configurations. Specifically, we use three window sizes: $[1024, 2048, 512]$, corresponding hop sizes: $[160, 240, 50]$, and window lengths: $[600, 1200, 240]$. These multi-resolution settings enable a robust evaluation of the generated audio’s structure across different temporal and spectral granularities.

\vspace{0.3em}

This SC-based analysis is particularly insightful for latent diffusion models (LDMs), which typically employ a \textit{VAE encoder-decoder pipeline} and/or \textit{neural vocoders} to reconstruct waveform outputs. The SC score thus helps us identify how much information is lost or distorted during the decoding process, making it a critical metric for analyzing the fidelity of waveform reconstruction.

\subsection{results}

\begin{table}[ht]
\centering
\caption{Tone Morphing Quality Evaluation across Diverse Methods and LDM Variants with Different LoRA Configurations. $\downarrow$ Indicates Lower is Better; $\uparrow$ Indicates Higher is Better.}
\label{tab:interpolation_metrics}
\scriptsize  
\setlength{\tabcolsep}{5pt}  
\renewcommand{\arraystretch}{1.1}  
\begin{tabular}{|l|c|c|}
\hline
\textbf{Model} & \textbf{CDPAM$_\text{mean} \pm \text{std}$ $\downarrow$} & \textbf{MOS $\uparrow$} \\
\hline
\multicolumn{3}{|c|}{\textbf{AudioLDM}} \\
\hline
w/o LoRA  & 0.32 ± 0.100 & \textbf{3.17} \\
w/ 1 LoRA & 0.45 ± 0.140 & 1.07 \\
w/ 2 LoRA & \textbf{0.22 ± 0.132} & 3.03 \\
\hline
\multicolumn{3}{|c|}{\textbf{AudioLDM2}} \\
\hline
w/o LoRA  & \textbf{0.25 ± 0.120} & \textbf{3.30} \\
w/ 1 LoRA & 0.34 ± 0.110 & 2.70 \\
w/ 2 LoRA & 0.33 ± 0.122 & 3.23 \\
\hline
\multicolumn{3}{|c|}{\textbf{MusicLDM}} \\
\hline
w/o LoRA  & 0.85 ± 0.116 & 1.97 \\
w/ 1 LoRA & 0.19 ± 0.120 & \textbf{3.70} \\
w/ 2 LoRA & \textbf{0.08 ± 0.114} & 1.20 \\
\hline
\multicolumn{3}{|c|}{\textbf{Spherical Music2Latent}} \\
\hline
Interpolation & \textbf{0.13 ± 0.060} & \textbf{4.3} \\
\hline
\end{tabular}
\end{table}

\begin{table}[ht]
\centering
\caption{Spectral Convergence (SC) Loss across Different Vocoder Frameworks, with and without VAE. A Lower SC means Better Performance.}
\label{tab:sc_loss}
\begin{tabular}{|l|c|c|}
\hline
\textbf{Model} & \textbf{Mean} & \textbf{Median} \\
\hline
\multicolumn{3}{|c|}{\textbf{Without VAE}} \\
\hline
BigVGAN & \textbf{0.03978} & \textbf{0.04005} \\
HifiGAN (AudioLDM) & 0.72955 & 0.15795 \\
HifiGAN (AudioLDM2) & 0.72955 & 0.15795 \\
HifiGAN (MusicLDM) & 0.30206 & 0.15017 \\
\hline
\multicolumn{3}{|c|}{\textbf{With VAE}} \\
\hline
HifiGAN (AudioLDM) & 1.80133 & 0.79835 \\
HifiGAN (AudioLDM2) & 0.49473 & 0.13271 \\
HifiGAN (MusicLDM) & \textbf{0.46880} & \textbf{0.11239} \\
BigVGAN (AudioLDM) & 1.10628 & 0.35980 \\
BigVGAN (AudioLDM2) & 0.50800 & 0.12355 \\
BigVGAN (MusicLDM) & 0.65704 & 0.22113 \\
\hline
\end{tabular}
\end{table}

\paragraph{Perceptual Quality Evaluation}

We evaluated the performance of various models on the guitar tone morphing task, as summarized in \textbf{Table~\ref{tab:interpolation_metrics}}. The assessment focused on two key metrics: \textbf{CDPAM mean ± std} and \textbf{Mean Opinion Score (MOS)}. Among all systems, spherical interpolation using Music2Latent achieved the highest MOS, indicating that listeners consistently rated its outputs as the most natural and musically coherent.

What stands out is the performance of spherical interpolation using Music2Latent, which achieved the best subjective rating despite being structurally different from other LDMs. Unlike other models that rely on VAE and vocoder components, it performs interpolation directly in the latent space and uses a different decoding strategy. This may be helpful for preserving structural and stylistic continuity between source and target tones, contributing to its superior perceptual quality.

In conclusion, our results highlight that \textbf{structure-preserving latent interpolation} can play a crucial role in improving the subjective quality of tone morphing. The following section explores how VAE and vocoder components can affect reconstruction quality in LDM.

\paragraph{Reconstruction Capabilities of Different Frameworks}

We evaluated the reconstruction performance of different structures using the training dataset. For models without a VAE, the audio is first converted to a Mel spectrogram and then directly reconstructed using a vocoder. For models that include a VAE, the spectrogram is first encoded and decoded through the VAE before being passed to the vocoder for waveform generation.

\textbf{BigVGAN}\cite{lee2022bigvgan} uses a consistent pre-trained model, \texttt{nvidia/bigvgan\_v2\_44khz\_128band\_512x}. For HiFiGAN-based models, we also apply the same set of pre-trained checkpoints to ensure fair comparison. The SC loss results across different setups are shown in Table\ref{tab:sc_loss}.

From the table, we can see that BigVGAN performs best in terms of SC loss when used without a VAE. However, when paired with a VAE trained within the HiFiGAN framework, reconstruction quality improves further. This suggests a strong compatibility between the VAE and vocoder in that setup. However, combining BigVGAN with VAEs from other LDM architectures results in higher SC loss, possibly due to a mismatched design.

These findings highlight that many existing models rely on several interdependent components, such as VAEs, vocoders, and auxiliary modules, to achieve good audio quality. This reliance adds complexity, risks error propagation, and destabilizes reconstruction. In contrast, our proposed model removes these dependencies by design, offering a simpler pipeline that is easier to train and more robust. This streamlined structure ensures stable performance while maintaining competitive reconstruction fidelity.
\section{Conclusion}

Our study highlights the critical importance of an effective encoder in achieving high-quality audio generation and reconstruction. Extensive experiments demonstrated that the proposed architecture of spherical interpolation using Music2Latent outperforms traditional VAE-based models, delivering superior audio clarity and fidelity. The enhanced encoder design enables more precise latent space representations, directly contributing to improve the output quality.

A key novelty of this work is to apply dual-sided LoRA fine-tuning on the latent diffusion model. By fine-tuning both the forward and backward processes within the U-Net structure, we achieve greater stability and consistency in the generated audio. Experiments show that this approach significantly improves robustness across diverse inputs.

Looking forward, the strong reconstruction capabilities of the proposed model suggest promising extensions. One possible extension is local tone control within music, enabling users to adjust specific segments or instruments in a track with fine granularity. Another one is to leverage its latent space for advanced source separation, including isolating vocals, drums, or other components with minimal artifacts, leading to new possibilities in remixing and music production.

\printbibliography

\end{document}